\documentclass[
preprint,
showpacs,%preprintnumbers,
nobibnotes,
 amsmath,amssymb,
 aps,
pra,
floatfix,
]{revtex4-1}

\usepackage{latexsym} 
\usepackage{enumerate}
\usepackage{bm}% bold math
\usepackage{graphicx}% Include figure files
\usepackage[usenames]{color} 
\usepackage{dcolumn}% Align table columns on decimal point
\begin{document}

%\preprint{APS/123-QED}

\title{Classification of Spin-Nematic Squeezing in Spin-1 Collective Atomic Systems}% Force line breaks with \\

\author{Emi Yukawa}
 \affiliation{National Institute for Informatics, 2-1-2 Hitotsubashi, Chiyoda-ku, Tokyo 101-8430, Japan}%Lines break automatically or can be forced with \\
%\affiliation{%


\author{Masahito Ueda}
\affiliation{
Department of Physics, University of Tokyo, 7-3-1 Hongo, Bunkyo-ku, Tokyo 113-0033, Japan 
}%
\author{Kae Nemoto}
\affiliation{%
 National Institute for Informatics, 2-1-2 Hitotsubashi, Chiyoda-ku, Tokyo 101-8430, Japan 
}%

\date{\today}% It is always \today, today,

\begin{abstract}
In spin-1 collective atomic systems, the spin and nematic-tensor operators constitute the su(3) Lie algebra whose su(2) subalgebras are shown to give 
two distinct classes of squeezing which are unitarily equivalent to spin squeezing and spin-nematic squeezing. 
We explicitly construct a unitary operator that generates  an arbitrary squeezed spin-nematic state from an arbitrary Fock state. 
In particular, we demonstrate that squeezed spin states can be generated from a polar state and that squeezed spin-nematic sates can be generated 
from a fully spin-polarized state. 

%\begin{description}
% Secondary publications and information retrieval purposes.
%03.75.Kk, 03.75.Mn, 05.30.Jp
% You may use the \texttt{description} environment to structure your abstract;
% use the optional argument of the \verb+\item+ command to give the category of each item. 
%\end{description}
\end{abstract}

\maketitle

%\tableofcontents

\section{\label{sec:1}Introduction}
Squeezing in spin systems~\cite{KU,Wine} has been an active research field in atomic, molecular, and optical physics. 
Squeezing is one of the most promising candidates in precision measurement to go beyond the standard quantum limit (SQL)~\cite{Wine,Maccone,DalRev,Nori,Klempt,GrossRev}. 
It has been suggested that spin squeezing can be utilized for magnetometers~\cite{Mabuchi1,Mabuchi2,Molmer1,Stamper-Kurn1,Stamper-Kurn2,Thompson} and 
atomic clocks~\cite{Wine,Polzik1,Oates,Ye}, while squeezed spin states have been demonstrated  in several physical systems~\cite{Polzik3,Bigelow,Kasevich,Polzik4,Mabuchi1,Polzik5,Oberthaler1,Vuletic,Oberthaler2,Treutlein,Thompson,Chapman,Takahashi2}.   

A squeezed spin state (SSS) can be generated from a coherent spin state (CSS)~\cite{Radcliffe,Arecchi} in which $N$ pseudo spin-1/2 particles point the same direction 
with spin fluctuations perpendicular to the average spin direction being completely random. 
Therefore, the magnitude of the spin fluctuations of a CSS is proportional to $\sqrt{N}$, which defines the standard quantum limit (SQL) of spin systems. 
This SQL can be surpassed by spin squeezing in which spin fluctuations in one direction are reduced at the expense of enhanced fluctuations in the 
other direction by making individual spin fluctuations quantum-mechanically correlated via nonlinear interactions~\cite{KU,DalRev,GrossRev,Mabuchi1,Mabuchi2,Thompson,Collet,Zoller1,Oberthaler1,Oberthaler2,Treutlein,Bigelow,Vuletic,Lukin,Takahashi1,Takahashi2,LiYou1} such as one-axis twisting and two-axis counter twisting.

Bose-Einstein condensates (BECs) of ultracold atomic gases with internal degrees of freedom, i.e., spinor BECs, are among the ideal candidates to 
achieve spin squeezing.  
BECs can naturally be prepared in any desired CSS and versatile tools for introducing nonlinear interactions necessary for spin squeezing are available. 
One-axis twisting can be realized by the Josephson coupling between two condensate modes~\cite{DalRev,GrossRev,Collet,Zoller1,Oberthaler1,Oberthaler2,Treutlein} and atom-field interactions~\cite{Mabuchi1,Mabuchi2,Thompson,Bigelow,Vuletic,Takahashi1,Takahashi2}. 
In the case of pseudo spin-$1/2$ BECs, squeezing has been extensively investigated by utilizing two appropriate magnetic sublevels or two-mode condensates~\cite{DalRev,GrossRev,Collet,Molmer2,Zoller1,Oberthaler1,Oberthaler2,Treutlein}. 
In fact, quantum fluctuations below the SQL have been observed in two-mode BECs with the Josephson-type one-axis twisting~\cite{Oberthaler2,Treutlein}. 

In the case of pseudo spin-$1/2$ BECs, spin squeezing is the only possibility for reducing quantum fluctuations since the square of any Pauli 
matrix is equal to the identity operator and the anti-commutator of any two different Pauli matrices vanishes. 
Therefore, only inter-spin correlations can be utilized for suppressing quantum fluctuations. 
However, other types of squeezing are possible for higher spins. 
A systematic investigation of squeezing for the case of spin-1 collective atomic systems is the primary purpose of this paper. 

In the case of collective spin-1 systems, the square of any spin-1 matrix $J_{\mu}$ ($\mu = x,y,z$) and the anti-commutator of two different spin matrices are not proportional to the identity 
matrix but represent the quadrupolar moments, namely, the nematic tensor (or alignment tensor~\cite{Polzik2,Polzik3}), and the other multipolar moments can be expressed in terms 
of the spin vector $\vec{\bm{J}} = {(J_x,J_y,J_z)}^T$ and the nematic tensor. 
The spin and nematic-tensor operators constitute the su(3) Lie algebra, which suggests a new trade-off relations between spin and nematic-tensor operators.
This implies that we can control not only quantum fluctuations of the spin vector but also those of the nematic tensor by manipulating various types of correlations between 
non-commutative spin and nematic-tensor observables. 
Here, we refer to squeezing on any two components of the spin vector as spin squeezing and that on any two components of the spin vector and the nematic 
tensor as spin-nematic squeezing~\cite{Polzik2,Polzik3,Stamper-Kurn2,Chapman}. 
The spin-nematic squeezing was first realized in Ref.~\cite{Polzik3}, which was recognized as spin squeezing in the spin and the alignment tensor, that is, the nematic tensor. 
Recently, spin-nematic squeezing has experimentally surpassed 8dB below the SQL~\cite{Chapman}. 
Such a squeezing can be interpreted as two-mode squeezing involving two of the three magnetic sublevels of each spin-1 particle~\cite{Meystre,Zoller2,LiYou2}. 
The spin-nematic squeezing is also referred to as $U$-$V$ spin squeezing and isospin squeezing within the SU(3) framework~\cite{LiYou2}. 
Squeezing induced by other types of correlations such as inter-nematic ones should also be possible. 
 
In the present paper, we address the questions of what observables and to what extent we can squeeze beyond the SQL for spin-1 BECs. 
We will show that there exist two qualitatively different (type-1 and type-2) classes of squeezing in spin-1 BECs based on their underlying Lie algebras. 
Type-1 squeezing involves spin squeezing, while type-2 does spin-nematic squeezing, implying that type-1 squeezing and type-2 squeezing 
are unitarily equivalent to spin squeezing and spin-nematic squeezing, respectively. 
As we see later in details, this leads us to a counterintuitive conclusion: we may generate both type-1 squeezing and type-2 squeezing from any 
CSS. 
A fully-polarized CSS with no nematicity can be spin-nematically squeezed and a polar CSS with zero averaged spin can be spin squeezed. 
We will also demonstrate that both type-1 squeezing and type-2 squeezing can be generated through one-axis twisting and discuss their squeezing limits. 

This paper is organized as follows. 
In Sec.~\ref{sec:2}, we classify squeezing in spin-1 BECs into two types, type 1 and type 2, based on the underlying Lie algebra. 
In Sec.~\ref{sec:3}, we show that the type-1 and type-2 squeezed states can be generated through one-axis twisting and discuss their squeezing 
limits.  In Sec.~\ref{sec:4}, we summarize the main results of this paper. 
Some complicated Lie-algebraic manipulations are discussed in Appendices to avoid digressing from the main subjects.
 
\section{\label{sec:2}Lie-Algebraic Classification of Squeezing in Spin-1 BECs} 
\subsection{\label{ssec:2-1}Second-Quantized Spin and Nematic-Tensor Operators} 
We consider a BEC consisting of $N$ identical spin-1 bosons and take a single-mode approximation to the BEC, that is, we assume 
that all the bosons share the same spatial mode. 
Then, the state of the system can be described by space independent one-particle creation and annihilation operators ${\hat{a}}_m^{\dagger}$ and 
${\hat{a}}_m$, where $m$ indicates the magnetic sublevels $m=-1,0,1$. 
The second-quantized forms of the spin vector ${\hat{J}}_{\mu}$ and the nematic tensor ${\hat{N}}_{\mu \nu}$ ($\mu ,\nu = x,y,z$) in the basis of the magnetic sublevels are expressed as follows: 
\begin{equation}
	{\hat{J}}_{\mu} = \sum_{m,n=-1,0,1} {\left (J_{\mu} \right )}_{mn} {\hat{a}}_m^{\dagger} {\hat{a}}_n, \label{eq:1-1} 
\end{equation} 
\begin{equation} 
	{\hat{N}}_{\mu \nu} = \frac{1}{2}  \sum_{m,n=-1,0,1} {\left ( J_{\mu} J_{\nu} + J_{\nu} J_{\mu} \right )}_{mn} {\hat{a}}_m^{\dagger} {\hat{a}}_n. 
	\label{eq:1-2}
\end{equation} 
Here, $J_{\mu}$ expresses the spin matrix for the spin quantum number $J=1$. 
As mentioned in the previous section, quantum fluctuations can be redistributed via squeezing between two non-commutative observables which can be expressed as linear combinations of ${\hat{J}}_{\mu}$ and ${\hat{N}}_{\mu \nu}$. 
However, the components of the spin vector and the nematic tensor are not, in general, linearly independent. 
Among them, we can choose the following eight linearly-independent observables as generators of the su(3) Lie algebra: 
\begin{equation}
	\{ {\hat{J}}_x, {\hat{J}}_y, {\hat{J}}_z, {\hat{Q}}_{xy}, {\hat{Q}}_{yz}, {\hat{Q}}_{zx}, {\hat{D}}_{xy}, \hat{Y} \} \equiv  \{ {\hat{\Lambda}}_i \}, \ 
	(i =1,\cdots ,8), \label{eq:1-3}	
\end{equation}
where ${\hat{\Lambda}}_1={\hat{J}}_x$, ${\hat{\Lambda}}_2={\hat{J}}_y$, ..., ${\hat{\Lambda_8}}=\hat{Y}$, and
\begin{align} 
	&{\hat{Q}}_{\mu \nu} \equiv 2{\hat{N}}_{\mu \nu}, \label{eq:1-4} \\
	&{\hat{D}}_{xy} \equiv {\hat{N}}_{xx} - {\hat{N}}_{yy}, \label{eq:1-5} \\
	&\hat{Y} \equiv \frac{1}{\sqrt{3}} \left ( - {\hat{N}}_{xx} - {\hat{N}}_{yy} + 2 {\hat{N}}_{zz} \right ). \label{eq:1-6} 
\end{align}
The matrix representations of these observables are given by 
\begin{equation} 
\begin{split}
	&J_x = \frac{1}{\sqrt{2}} \begin{pmatrix} 0 & 1 & 0 \\ 1 & 0 & 1 \\ 0 & 1 & 0 \end{pmatrix}, \ 
	J_y = \frac{i}{\sqrt{2}} \begin{pmatrix} 0 & -1 & 0 \\ 1 & 0 & -1 \\ 0 & 1 & 0 \end{pmatrix}, \ 
	J_z = \begin{pmatrix} 1 & 0 & 0 \\ 0 & 0 & 0 \\ 0 & 0 & -1 \end{pmatrix}, \\
	&Q_{xy} = i \begin{pmatrix} 0 & 0 & -1 \\ 0 & 0 & 0 \\ 1 & 0 & 0 \end{pmatrix}, \ 
	Q_{yz} = \frac{i}{\sqrt{2}} \begin{pmatrix} 0 & -1 & 0 \\ 1 & 0 & 1 \\ 0 & -1 & 0 \end{pmatrix}, \ 
	Q_{zx} = \frac{1}{\sqrt{2}} \begin{pmatrix} 0 & 1 & 0 \\ 1 & 0 & -1 \\ 0 & -1 & 0 \end{pmatrix}, \\
	&D_{xy} = \begin{pmatrix} 0 & 0 & 1 \\ 0 & 0 & 0 \\ 1 & 0 & 0 \end{pmatrix}, \ 
	Y = \frac{1}{\sqrt{3}} \begin{pmatrix} 1 & 0 & 0 \\ 0 & -2 & 0 \\ 0 & 0 & 1 \end{pmatrix}. \label{eq:1-7}
\end{split}
\end{equation} 
Thus, an arbitrary observable $\hat{A}$ can be expressed in terms of ${\Lambda}_i$'s as  
\begin{equation} 
	\hat{A} = \sum_{i=1}^8 c_i {\hat{\Lambda}}_i, \ (c_i \in \mathbb{R}, \ \sum_{i=1}^8 c_i^2 = 1), \label{eq:1-8}
\end{equation} 
except for an overall normalization constant. 

\subsection{\label{ssec:2-2}Type-1 Squeezing and Type-2 Squeezing} 
Squeezing concerns the trade-off relation between non-commutative operators. 
In spin-1 BECs, such operators can be expressed as Eq.~(\ref{eq:1-8}) within the framework of the su(3) Lie algebra. 
The su(3) Lie algebra involves su(2) subalgebras whose triads of generators satisfy cyclic commutation relations such as $[ {\hat{J}}_x, {\hat{J}}_y ] = i{\hat{J}}_z$. 
These commutation relations lead to the trade-off relations in quantum fluctuations of the corresponding observables, and hence we can consider squeezing by choosing a triad of su(2) generators. 
Such an su(2)-type squeezing can be classified into two distinct classes since the su(3) Lie algebra involves two qualitatively different su(2) subalgebras 
that can be distinguished by their structure constants. 
Here, to classify su(2)-type squeezing, we define $\lambda$ in terms of the structure constant $f_{ij}^k$ of an su(2) subalgebra as follows: 
\begin{equation} 
\begin{split}
	&\lambda = \left | f_{ij}^k \right |, \\ &[ {\hat{X}}_i, {\hat{X}}_j ] = i f_{ij}^k {\hat{X}}_k, \label{eq:2-0} 
\end{split}
\end{equation}  
where observables ${\hat{X}}_i$, ${\hat{X}}_j$, and ${\hat{X}}_k$ are generators of the su(2) subalgebra. 

To identify all su(2) subalgebras in the su(3) Lie algebra, we first construct the root diagram of the su(3) Lie algebra as discussed in Appendix~\ref{as1}. 
The root diagram, which uniquely determines the structure of the Lie algebra, implies that the su(3) Lie algebra involves two types of su(2) subalgebras, where $\lambda$ defined in Eq.~(\ref{eq:2-0}) takes either on $\lambda =1$ or $\lambda=2$ (see Appendix~\ref{as1}). 
Here and henceforth, we refer to the su(2) subalgebras with $\lambda = 1$ as type 1 and those with $\lambda =2$ as type 2. 
Here, we note that two su(2) subalgebras $\{ {\hat{X}}_i, {\hat{X}}_j, {\hat{X}}_k \}$ and $\{ {\hat{X}}_i^{\prime}, {\hat{X}}_j^{\prime}, {\hat{X}}_k^{\prime} \}$ 
of the same type can transform to each other through an SU(3) rotation $\hat{U}$, i.e., 
\begin{equation} 
	\{ {\hat{X}}_i^{\prime}, {\hat{X}}_j^{\prime}, {\hat{X}}_k^{\prime} \} = \{ \hat{U} {\hat{X}}_i {\hat{U}}^{\dagger}, \hat{U} {\hat{X}}_j {\hat{U}}^{\dagger}, \hat{U} {\hat{X}}_k {\hat{U}}^{\dagger} \}, \label{eq:2-10} 
\end{equation} 
since the root diagram is independent of the basis of a particular representation, whose transformation can be expressed as an SU(3) rotation. 
This implies that all type-1 su(2) subalgebras are unitarily equivalent to the set of the spin components $\{ {\hat{J}}_x, {\hat{J}}_y, {\hat{J}}_z \}$. 
Thus, squeezing in triads of a type-1 subalgebra can be considered to be equivalent to spin squeezing. 
Similarly, any type-2 su(2) subalgebra can be transformed into a set of spin and nematic-tensor observables $\{ {\hat{D}}_{xy}, {\hat{Q}}_{xy}, {\hat{J}}_{z} \}$, in which 
spin-nematic squeezing has been observed~\cite{Polzik3}, which implies that squeezing in triads of a type-2 subalgebra can be considered to be equivalent to spin-nematic 
squeezing.
Here, the su(2) subalgebra $\{ {\hat{D}}_{xy}, {\hat{Q}}_{xy}, {\hat{J}}_{z} \}$ can also be transformed into $\{ {\hat{D}}_{yz}, {\hat{Q}}_{yz}, {\hat{J}}_{x} \}$ of spin-nematic 
squeezing in Ref.~\cite{Chapman}. 
We also note that type-2 squeezing can be interpreted as two-mode squeezing~\cite{Meystre,Zoller2,LiYou2}. 
Here, observables in a type-2 su(2) subalgebra can be shown to be unitarily equivalent to Pauli operators 
defined with respect to two-mode creation and annihilation operators since the type-2 su(2) subalgebra $\{ {\hat{D}}_{xy}, {\hat{Q}}_{xy}, {\hat{J}}_z \}$ 
can be expressed in terms of ${\hat{a}}_{\pm1}^{\dagger}$ and ${\hat{a}}_{\pm1}$. 

\section{\label{sec:3}Spin-Nematic Squeezing in Spin-1 BECs} 
Type-1 and type-2 squeezed states can be generated through one-axis twisting from a CSS. 
In the procedure of squeezing, we start with a BEC prepared in a CSS which is nothing but the Fock state defined as 
\begin{equation} 
	\left | {\Psi}_{\mathrm{Fock}} \right > = \frac{1}{\sqrt{N!}} {\hat{\mathfrak{a}}}^{\dagger N} \left | \mathrm{vac} \right >, \label{eq:1-0-1} 
\end{equation} 
where ${\hat{\mathfrak{a}}}^{\dagger}$ can be written in terms of creation operators with magnetic sublevels $m=-1,0,1$ as 
\begin{equation}
	{\hat{\mathfrak{a}}}^{\dagger} = {\zeta}_1 a_1^{\dagger} + {\zeta}_0 a_0^{\dagger} + {\zeta}_{-1} a_{-1}^{\dagger}, \ 
	\sum_{m=-1,0,1} {\left | {\zeta}_m \right |}^2 = 1. \label{eq:1-0-2} 
\end{equation} 
Both type-1 squeezing and type-2 squeezing can be generated from an arbitrary Fock state. 
In particular, we can produce spin squeezing starting with an unpolarized Fock state, although such a state has no spin expectation values 
$\langle {\bm{J}} \rangle = \bm{0}$. 
Similarly, spin-nematic squeezing can be generated from a fully-polarized Fock state which has no nematicity. 
In the following subsections, we show that type-1 and type-2 squeezed states are generated from a general Fock state in the case of spin-1 BECs and 
discuss the squeezing limit for each case. 

\subsection{\label{ssec:3-1}Type-1 Squeezing} 
First, let us assume that the initial state is given by Eq.~(\ref{eq:1-0-1}) and generate a type-1 spin squeezed state through one-axis twisting. 
We begin by noting that an SU(3) rotation ${\hat{\mathcal{U}}}_1 (\alpha ,\beta ,\gamma ,\varphi)$ 
\begin{equation}
	{\hat{\mathcal{U}}}_1 (\alpha ,\beta ,\gamma ,\varphi) 
	= \exp {(-i\alpha {\hat{J}}_z)} \exp {(-i\beta {\hat{J}}_y)} \exp {(-i\gamma {\hat{J}}_z)} \exp {(-i\varphi {\hat{Q}}_{yz})}, \label{eq:3-2} 
\end{equation} 
satisfies 
\begin{equation}
	\left | {\Psi}_{\mathrm{Fock}} \right > = \frac{1}{\sqrt{N!}} {\left ( {\hat{\mathcal{U}}}_1 (\alpha ,\beta ,\gamma ,\varphi ) e^{-i \frac{\pi}{2} {\hat{J}}_y} 	
	{\hat{a}}_1^{\dagger} e^{i \frac{\pi}{2} {\hat{J}}_y} {\hat{\mathcal{U}}}_1^{\dagger} (\alpha ,\beta ,\gamma ,\varphi ) \right )}^N 
	\left | \mathrm{vac} \right >. \label{eq:3-3} 
\end{equation} 
By rotating the set of observables of the su(2) subalgebra of type 1, $\{ {\hat{J}}_x , {\hat{J}}_y, {\hat{J}}_z \}$, through a unitary operator 
${\hat{\mathcal{U}}}_1$, we obtain a new set of observables which also serve as generators of a type-1 su(2) subalgebra: 
\begin{equation}
	\{ {\hat{J}}_x^{\prime} , {\hat{J}}_y^{\prime}, {\hat{J}}_z^{\prime} \} 
	\equiv \{ {\hat{\mathcal{U}}}_1 {\hat{J}}_x {\hat{\mathcal{U}}}_1^{\dagger}, {\hat{\mathcal{U}}}_1 {\hat{J}}_y {\hat{\mathcal{U}}}_1^{\dagger}, 
	{\hat{\mathcal{U}}}_1 {\hat{J}}_z {\hat{\mathcal{U}}}_1^{\dagger} \}. \label{eq:3-4} 
\end{equation} 
Then, the initial state $\left | {\Psi}_{\mathrm{Fock}} \right >$ becomes a state that is fully polarized along the ${\hat{J}}_x^{\prime}$ direction 
in the SU(2) sphere of radius $N$ with the three orthogonal axes given by $\{ {\hat{J}}_x^{\prime} , {\hat{J}}_y^{\prime}, {\hat{J}}_z^{\prime} \}$. 
%\begin{figure}[t]
%\centering
%\includegraphics[width=5cm,clip]{temp.eps}
%\caption{CSS1} 
%\label{fig:3-1}
%\end{figure} 
Thus, the one-axis twisting Hamiltonian for $\left | \mathrm{Fock} \right >$ in Eq.~(\ref{eq:1-0-1}) is expressed as 
\begin{equation} 
	{\hat{H}}_{\mathrm{one-axis}}^{\prime} = \hbar \chi {\hat{J}}_z^{\prime 2}, \label{eq:3-6} 
\end{equation} 
where $\chi$ denotes the strength of interaction.  
A type-1 squeezed state provides the same minimum variance as in the case of spin squeezing~\cite{KU}, because the magnitude of the structure constant 
in $\{ {\hat{J}}_x^{\prime} , {\hat{J}}_y^{\prime}, {\hat{J}}_z^{\prime} \}$ is the same as $\{ {\hat{J}}_x, {\hat{J}}_y, {\hat{J}}_z \}$. 
Provided the number of particles $N$ is so large and the time $t$ is so small that $N \chi t \gg 1$ and $N {(\chi t)}^2 \ll 1$, 
the variance is minimized at $\chi t={(3/8N^4)}^{1/6}$, giving 
\begin{equation}
	{\left <  {\Delta}^2 \right >}_{\mathrm{min}} \simeq \frac{1}{4} {\left ( \frac{9N}{4} \right )}^{1/3}, \label{eq:3-7} 
\end{equation} 
where $\Delta$ is given by 
\begin{equation} 
	\Delta = \sin {\nu} {\hat{J}}_y^{\prime} + \cos {\nu} {\hat{J}}_z^{\prime},  \ \nu \simeq \frac{1}{2} 
	\left [ \arctan {(N \chi t)} - \chi t \right ]. \label{eq:3-8} 
\end{equation} 

If the initial state is in a polar phase with $m=0$, for instance, the corresponding Fock state is given by 
$\left | {\Psi}_{\mathrm{Fock}} \right >_{\mathrm{polar}} = {(N!)}^{-1/2} {\hat{a}}^{\dagger N}_0 \left | \mathrm{vac} \right >$ and thus 
$(\alpha ,\beta ,\gamma ,\varphi ) = (0,0,0, -3\pi /4)$. 
The type-1 subalgebra, where $\left | {\Psi}_{\mathrm{Fock}} \right >_{\mathrm{polar}}$ is polarized in the $x$ direction, is obtained as 
$\{ {\hat{J}}_x^{\prime} ,{\hat{J}}_y^{\prime} ,{\hat{J}}_z^{\prime} \} = \{ {\hat{D}}_{yz}, (-{\hat{J}}_y+{\hat{Q}}_{xy})/\sqrt{2}, -({\hat{J}}_z+{\hat{Q}}_{zx})/\sqrt{2} \}$ 
and the one-axis twisting Hamiltonian becomes ${\hat{H}}_{\mathrm{one-axis}}^{\prime} = \hbar \chi {({\hat{J}}_z + {\hat{Q}}_{zx})}^2 /2$. 
The squeezing limit of one-axis twisting is given by Eq.~(\ref{eq:3-7}). 

\subsection{\label{ssec:3-2}Type-2 Squeezing} 
The type-2 squeezing can also be generated from the initial state given by Eq.~(\ref{eq:1-0-1}) with a method similar to the case of type-1 squeezed spin states. 
Let us consider an SU(3) rotation 
\begin{equation}
{\hat{\mathcal{U}}}_2 (\alpha ,\beta ,\gamma ,\varphi) 
	= \exp {(-i\alpha {\hat{J}}_z)} \exp {(-i\beta {\hat{J}}_y)} \exp {(-i\gamma {\hat{J}}_z)} \exp {(-i\varphi {\hat{Q}}_{xy})}, \label{eq:3-9-1}
\end{equation}
where $\alpha$, $\beta$, $\gamma$, and $\varphi$ satisfy 
\begin{equation} 
	\left | {\Psi}_{\mathrm{Fock}} \right > = \frac{1}{\sqrt{N!}} {\left ( {\hat{\mathcal{U}}}_2 (\alpha ,\beta ,\gamma ,\varphi +\pi /4)  
	{\hat{a}}_1^{\dagger} {\hat{\mathcal{U}}}_2^{\dagger} (\alpha ,\beta ,\gamma ,\varphi + \pi /4) \right )}^N 
	\left | \mathrm{vac} \right >, \label{eq:3-9-2} 
\end{equation} 
and rotate the su(2) subalgebra $\{ {\hat{D}}_{xy}, {\hat{Q}}_{xy}, {\hat{J}}_z \}$ through ${\hat{\mathcal{U}}}_2 (\alpha ,\beta ,\gamma ,\varphi )$ as 
\begin{equation} 
	\{ {\hat{D}}_{xy}^{\prime \prime}, {\hat{Q}}_{xy}^{\prime \prime}, {\hat{J}}_z^{\prime \prime} \} 
	= \{ {\hat{\mathcal{U}}}_2 {\hat{D}}_{xy} {\hat{\mathcal{U}}}_2^{\dagger}, {\hat{\mathcal{U}}}_2 {\hat{Q}}_{xy} {\hat{\mathcal{U}}}_2^{\dagger}, {\hat{\mathcal{U}}}_2 {\hat{J}}_z {\hat{\mathcal{U}}}_2^{\dagger} \}. 
	\label{eq:3-10}
\end{equation} 
Then, the state $\left | {\Psi}_{\mathrm{Fock}} \right >$ appears to be fully polarized in the direction of the ${\hat{Q}}_{xy}^{\prime \prime}$-axis in the quasi-probabilistic 
representation in $\{ {\hat{D}}_{xy}^{\prime \prime}, {\hat{Q}}_{xy}^{\prime \prime}, {\hat{J}}_z^{\prime \prime} \}$. 
The initial state $\left | {\Psi}_{\mathrm{Fock}} \right >$ evolves under the one-axis Hamiltonian ${\hat{H}}^{\prime \prime} = \hbar \chi {\hat{J}}_z^{\prime \prime 2}$ into 
a squeezed state. 
In this case, the spin and the interaction energy are effectively modified as $N\to N/2$ and $\chi \to 2\chi$ because the the magnitude of the structure 
constant is twice as large as that of the type-1 su(2) subalgebra. 
The variance is minimized at $t=\chi t={(6/N^4)}^{1/6}$ in the limit $N\chi t \gg 1$ and $N {(\chi t)}^2 \ll 1$, giving 
\begin{equation} 
	{\left <  {\Delta}^2 \right >}_{\mathrm{min}} = \frac{1}{2} {\left ( 9N \right )}^{1/3}, \label{eq:3-11} 
\end{equation} 
which is about four times larger than the result in Eq.~(\ref{eq:3-7}), where $\Delta$ represents 
\begin{equation} 
	\Delta = \sin {\nu} {\hat{Q}}_{xy}^{\prime \prime} + \cos {\nu}  {\hat{J}}_z^{\prime \prime}, \ \nu \simeq \frac{1}{2} 
	\left [ \arctan {(N \chi t)} - 2 \chi t \right ]. \label{eq:3-12} 
\end{equation} 

In the case of a fully-polarized Fock state $\left | {\Psi}_{\mathrm{Fock}} \right >_{\mathrm{ferro}} =  {(N!)}^{-1/2} {\hat{a}}^{\dagger N}_1 \left | \mathrm{vac} \right >$, 
for instance, $(\alpha ,\beta ,\gamma ,\varphi ) = (0,0,0,-\pi /4 )$ and 
$\{ {\hat{D}}_{xy}^{\prime \prime}, {\hat{Q}}_{xy}^{\prime \prime}, {\hat{J}}_z^{\prime \prime} \} = \{ {\hat{J}}_z, {\hat{Q}}_{xy}, -{\hat{D}}_{xy} \}$. 
The one-axis twisting Hamiltonian is given by ${\hat{H}}_{\mathrm{one-axis}}^{\prime \prime} = \hbar \chi {\hat{D}}_{xy}^2$. 
The squeezing limit is given by Eq.~(\ref{eq:3-11}). 
 
\section{\label{sec:4}Conclusion} 
In this paper, we have shown that squeezing in a triad of observables in spin-1 BECs that constitute an su(2) subalgebra can be categorized into two 
classes, type 1 and type 2 according to the underlying su(2) subalgebra. 
Type-1 su(2) subalgebras are characterized by the magnitude of the structure constants being equal to $\lambda =1$ and are unitarily equivalent to the set of the spin operators 
$\{ {\hat{J}}_x, {\hat{J}}_y, {\hat{J}}_z \}$, 
while type-2 su(2) subalgebras have the structure constants with their magnitude being equal to $\lambda =2$ and are unitarily equivalent to $\{ {\hat{D}}_{xy}, {\hat{Q}}_{xy}, {\hat{J}}_z \}$. 
Thus type-1 and type-2 su(2) subalgebras can be transformed into $\{ {\hat{J}}_x, {\hat{J}}_y, {\hat{J}}_z \}$ and $\{ {\hat{D}}_{xy}, {\hat{Q}}_{xy}, {\hat{J}}_z \}$ 
through SU(3) rotations, respectively. 
This implies that both type-1 and type-2 squeezed states can be generated from an arbitrary CSS (or a Fock state) through one-axis twisting. 
We have explicitly demonstrated how to generate a squeezed spin state and a squeezed spin-nematic state from a polar Fock state and a 
fully-polarized Fock state, respectively. 
In the case of one-axis twisting, the squeezing limits for both types of squeezing are proportional to $N^{1/3}$with different proportionality coefficients due to different magnitudes of the structure constants. 

\appendix 
\section{\label{as1}Two su(2) subalgebras of the su(3) Lie Algebra} 
\begin{figure}[t]
\centering
\includegraphics[width=5cm,clip]{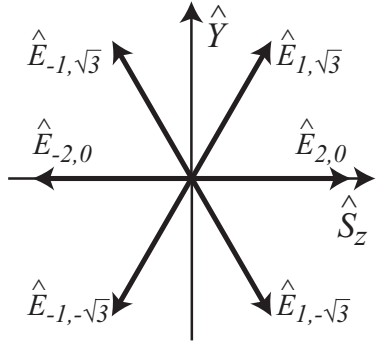}
\caption{Root diagram of the su(3) Lie algebra} 
\label{fig:2-1}
\end{figure}

The identification of the su(2) subalgebras from the su(3) Lie algebra can be done most straightforwardly in terms of the root diagram.
The root diagram of the su(3) Lie algebra is constructed as follows. 
A root diagram consists of root vectors given by sets of eigenvalues of adjoint representations of generators in a Cartan subalgebra. 
A Cartan subalgebra, which is a set of maximally Abelian operators among the generators of a Lie algebra, has two elements in the case of the su(3) Lie algebra. 
Here we set ${\hat{\Lambda}}_3 = {\hat{J}}_z$ and ${\hat{\Lambda}}_8 = \hat{Y}$ defined in Eqs.~(\ref{eq:1-6}) and (\ref{eq:1-7}) as elements of a Cartan subalgebra 
among the generators $\{ {\hat{\Lambda}}_i \}$ in Eq.~(\ref{eq:1-3}). 
The adjoint representations of ${\hat{J}}_z$ and $\hat{Y}$ are given by ${\{ \mathrm{ad}({\hat{\Lambda}}_3) \} }^j_i= f_{3i}^j$ and 
${\{ \mathrm{ad}({\hat{\Lambda}}_8) \} }^j_i=f_{8i}^j$ in terms of the structure constants $f_{ij}^k$'s defined as follows: 
\begin{equation}
	[ {\hat{\Lambda}}_i, {\hat{\Lambda}}_j ] = i \sum_{k=1}^8 f_{ij}^k {\hat{\Lambda}}_k. \label{eq:2-1}
\end{equation} 
The matrices  $\mathrm{ad}({\hat{\Lambda}}_3)$ and $\mathrm{ad}({\hat{\Lambda}}_8)$ can be simultaneously diagonalized to give simultaneous 
eigen operators ${\hat{E}}_{\bm{\alpha}}$'s accompanied by the root vectors $\bm{\alpha} = {({\alpha}_1,{\alpha}_2)}^T$'s, where 
${\alpha}_1$ and ${\alpha}_2$ express the eigenvalues of $\mathrm{ad}({\hat{\Lambda}}_3)$ and $\mathrm{ad}({\hat{\Lambda}}_8)$, respectively. 
The non-zero root vectors and their corresponding eigenoperators are given by 
\begin{align}
	& \begin{pmatrix} \pm 2 \\ 0 \end{pmatrix}, \ {\hat{E}}_{\pm 2,0} \equiv \frac{1}{\sqrt{2}} \left ( {\hat{D}}_{xy} \pm i{\hat{Q}}_{xy} \right ), \label{eq:2-2} \\
	& \begin{pmatrix} 1 \\ \pm \sqrt{3} \end{pmatrix}, \ 
	{\hat{E}}_{1,\pm \sqrt{3}} \equiv \frac{1}{2} \left [ {\hat{J}}_x \pm {\hat{Q}}_{zx} +  i \left ( {\hat{J}}_y \pm {\hat{Q}}_{yz} \right ) \right ], \label{eq:2-3} \\
	& \begin{pmatrix} -1 \\ \pm \sqrt{3} \end{pmatrix}, \ 
	{\hat{E}}_{-1,\pm \sqrt{3}} \equiv \frac{1}{2} \left [ {\hat{J}}_x \mp {\hat{Q}}_{zx} - i \left ( {\hat{J}}_{y} \mp {\hat{Q}}_{yz} \right ) \right ]. \label{eq:2-4} 
\end{align} 
Here we note that the eigen operators corresponding to the root vectors depend on the choice of a basis or a Cartan subalgebra, whereas the root vectors do not. 
The root diagram of the su(3) algebra involves the above six root vectors which form a hexagon as shown in Fig.~\ref{fig:2-1}. 
The root vectors can be expressed as linear combinations of the two simple roots, ${(1, \mp \sqrt{3})}^T \equiv {\bm{\alpha}}^{(1)}, {\bm{\alpha}}^{(2)}$. 

Next, we give all su(2) subalgebras that can be derived from the root diagram in Fig.~\ref{fig:2-1} and show that each of them can be categorized into one of the two classes 
referred to as type 1 and type 2 according to the magnitude of the structure constant $\lambda =1$ or $\lambda =2$. 
Three type-2 su(2) subalgebras can immediately be found in the root diagram in Fig.~\ref{fig:2-1}. 
Their raising and lowering operators are given by ${\hat{E}}_{\pm \bm{\alpha}}$'s corresponding to the pairs of the opposite root vectors, that is, 
${\hat{E}}_{\pm 2,0}$, ${\hat{E}}_{\pm 1, \pm \sqrt{3}}$, and ${\hat{E}}_{\pm 1, \mp \sqrt{3}}$. 

The su(2) subalgebras can be derived from them as follows. 
First, a pair ${\hat{E}}_{\pm \bm{\alpha}}$ and their associated root vectors $\pm \bm{\alpha}$ satisfy the following commutation relations: 
\begin{equation} 
	[ {\hat{J}}_z, {\hat{E}}_{\pm \bm{\alpha}} ] = \pm {\alpha}_1 {\hat{E}}_{\pm \bm{\alpha}}, \ 
	[ \hat{Y}, {\hat{E}}_{\pm \bm{\alpha}} ] = \pm {\alpha}_2 {\hat{E}}_{\pm \bm{\alpha}}, \ 
	[ {\hat{E}}_{\bm{\alpha}}, {\hat{E}}_{-\bm{\alpha}} ] = \bm{\alpha} \cdot \hat{\bm{H}}, \label{eq:2-5}
\end{equation} 
where $\hat{\bm{H}} \equiv {({\hat{J}}_z, \hat{Y})}^T$ and $\bm{\alpha} \cdot \hat{\bm{H}} \equiv {\alpha}_1 {\hat{J}}_z + {\alpha}_2 \hat{Y}$. 
Then, provided we take ${\hat{E}}_{\pm 2,0}$ in Eq.~(\ref{eq:2-2}) as an example, the equations in Eqs.~(\ref{eq:2-5}) become 
\begin{equation} 
	[ {\hat{J}}_z,  {\hat{D}}_{xy} \pm i {\hat{Q}}_{xy} ] = \pm 2 ( {\hat{D}}_{xy} \pm i {\hat{Q}}_{xy} ), \ 
	[ {\hat{D}}_{xy} + i {\hat{Q}}_{xy}, {\hat{D}}_{xy} - i {\hat{Q}}_{xy} ] = 4 {\hat{J}}_z, \label{eq:2-6}
\end{equation} 
which constitute the su(2) subalgebra $\{ {\hat{D}}_{xy}, {\hat{Q}}_{xy}, {\hat{J}}_z \}$ with $\lambda = 2$. 
Similarly to the case of ${\hat{E}}_{\pm 2,0}$, the remaining pairs ${\hat{E}}_{\pm 1, \pm \sqrt{3}}$ and ${\hat{E}}_{\pm 1, \mp \sqrt{3}}$ give the su(2) subalgebras 
$\{ ({\hat{J}}_x \pm {\hat{Q}}_{zx})/\sqrt{2}, ({\hat{J}}_y \pm {\hat{Q}}_{yz})/\sqrt{2}, ({\hat{J}}_z \pm \sqrt{3} \hat{Y})/2 \}$ with $\lambda = 2$. 
These two su(2) subalgebras are unitarily equivalent to $\{ {\hat{D}}_{xy}, {\hat{Q}}_{xy}, {\hat{J}}_z \}$ due to the hexagonal symmetry of the root diagram in 
Fig.~\ref{fig:2-1}. 
In fact, they can be transformed into $\{ {\hat{D}}_{xy}, {\hat{Q}}_{xy}, {\hat{J}}_z \}$ through a rotation about the ${\hat{Q}}_{xy}$-axis, 
${\hat{U}}_Q (\varphi ) \equiv \exp {(-i\varphi {\hat{Q}}_{xy}})$, and through the change of the basis, i.e., 
\begin{equation} 
\begin{split}
	&\{ ({\hat{J}}_x \pm {\hat{Q}}_{zx})/\sqrt{2}, ({\hat{J}}_y \pm {\hat{Q}}_{yz})/\sqrt{2}, ({\hat{J}}_z \pm \sqrt{3} \hat{Y})/2 \} \\
	&\xrightarrow{\{ \hat{A} \} \to \{ {\hat{U}}_Q (\pi/4) \hat{A} {\hat{U}}_Q^{\dagger}  (\pi/4) \} }
	\{ {\hat{J}}_x, {\hat{Q}}_{yz}, {\hat{D}}_{yz} \}, \ \{ -{\hat{Q}}_{zx}, {\hat{J}}_y, {\hat{D}}_{zx} \} \\ 
	&\xrightarrow{\text{change the basis}} \{ {\hat{D}}_{xy}, {\hat{Q}}_{xy}, {\hat{J}}_z \}. \label{eq:2-7}
\end{split}
\end{equation} 

On the other hand, the type-1 su(2) subalgebra can be obtained as follows. 
The four eigen operators, ${\hat{E}}_{1,\pm \sqrt{3}}$ and ${\hat{E}}_{-1,\pm \sqrt{3}}$, for instance, constitute the type-1 su(2) subalgebra. 
In this case, two raising operators are given by 
\begin{equation}
	{\hat{E}}_{1,\sqrt{3}} \pm {\hat{E}}_{1,-\sqrt{3}} = {\hat{J}}_x + i {\hat{J}}_y, \ {\hat{Q}}_{zx} + i {\hat{Q}}_{yz}, \label{eq:2-8} 
\end{equation}
and the corresponding lowering operators are given by ${\hat{E}}_{-1,-\sqrt{3}} \pm {\hat{E}}_{-1,\sqrt{3}}$. 
These raising and lowering operators provide the su(2) sublagebras $\{ {\hat{J}}_x, {\hat{J}}_y, {\hat{J}}_z \}$ and 
$\{ {\hat{Q}}_{zx}, {\hat{Q}}_{yz}, {\hat{J}}_z \}$ with $\lambda = 1$, which can be confirmed by the vanishing commutation relation between 
two eigen operators ${\hat{E}}_{\bm{\alpha}}$ and ${\hat{E}}_{\bm{\beta}}$, 
\begin{equation}
	[{\hat{E}}_{\bm{\alpha}},{\hat{E}}_{\bm{\beta}}] = 0, \label{eq:2-9} 
\end{equation} 
if and only if $\bm{\alpha} + \bm{\beta}$ does not coincides with a root vector. 
Here, $\{ {\hat{Q}}_{zx}, {\hat{Q}}_{yz}, {\hat{J}}_z \}$ is unitarily equivalent to $\{ {\hat{J}}_x, {\hat{J}}_y, {\hat{J}}_z \}$ through an SU(3) rotation 
about the ${\hat{Q}}_{xy}$ axis. 
Similarly to this case, the sets of four eigen operators $\{ {\hat{E}}_{\mp 1, \pm \sqrt{3}}, {\hat{E}}_{\pm 2,0} \}$ and 
$\{ {\hat{E}}_{\mp 2,0}, {\hat{E}}_{\pm 1, \pm \sqrt{3}} \}$ also provide the type-1 su(2) subalgebras, which are unitarily equivalent to the spin components 
$\{ {\hat{J}}_x, {\hat{J}}_y, {\hat{J}}_z \}$. 
It is shown that there exists no other class of su(2) subalgebra in Appendix~\ref{as2} for the sake of the completeness. 
As a consequence of the uniqueness of the root diagram, any su(2) subalgebra in the su(3) Lie algebra can be classified into either type 1 or type 2 
according to the magnitude of the structure constant $\lambda$ and two su(2) subalgebras of the same type can be transformed into each other 
through an SU(3) rotation, whereas those of different types cannot. 

\section{\label{as2} Number of Classes of su(2) Subalgebras in the su(3) Lie Algebra} 
We prove that there are no other classes of su(2) subalgebras in the su(3) Lie algebra than type 1 and type 2. 
In general, a raising operator of an su(2) subalgebra can be expressed in terms of the eigen operators corresponding to the root vectors as follows: 
\begin{equation} 
	{\hat{E}}_+ = \sum_{\bm{\alpha}} c_{\bm{\alpha}} {\hat{E}}_{\bm{\alpha}}, \label{eq:a-1}
\end{equation} 
where the sum of $\bm{\alpha}$ runs over all root vectors in Fig.~\ref{fig:2-1} and $c_{\bm{\alpha}}$'s are real. 
Here, without loss of generality, we set $c_{1,\sqrt{3}} = 1$ and $c_{-1,-\sqrt{3}} = 0$. 
The lowering operator associated with Eq.~(\ref{eq:a-1}) is given by ${\hat{E}}_- = {\hat{E}}_+^{\dagger}$. 
We obtain the commutation relation between these raising and lowering operators as follows: 
\begin{equation} 
\begin{split}
	[ {\hat{E}}_+, {\hat{E}}_- ] 
	&= [ {\hat{E}}_{1,\sqrt{3}}, {\hat{E}}_{-1,-\sqrt{3}} ] 
	+ ( c_{2,0}^2 - c_{-2,0}^2 ) [ {\hat{E}}_{2,0}, {\hat{E}}_{-2,0} ] 
	+ ( c_{1,-\sqrt{3}}^2 - c_{-1,\sqrt{3}}^2 ) [ {\hat{E}}_{1,-\sqrt{3}}, {\hat{E}}_{-1,\sqrt{3}} ] \\
	&+ \sqrt{2} \left [ c_{-1,\sqrt{3}} {\hat{E}}_{2,0} 
	+ ( c_{2,0} c_{1,-\sqrt{3}} - c_{-2,0} c_{-1,\sqrt{3}} ) {\hat{E}}_{1,\sqrt{3}} 
	- c_{2,0} {\hat{E}}_{1,-\sqrt{3}} + h.c. \right ], \label{eq:a-2}
\end{split}
\end{equation}
where we use the relations in Eq.~(\ref{eq:2-9}) and 
\begin{equation} 
\begin{split}
	&[ {\hat{E}}_{1,\sqrt{3}}, {\hat{E}}_{1,-\sqrt{3}}] = \sqrt{2} {\hat{E}}_{2,0}, \ [ {\hat{E}}_{2,0}, {\hat{E}}_{-1,-\sqrt{3}}] = - \sqrt{2} {\hat{E}}_{1,-\sqrt{3}}, \\ 
	&[ {\hat{E}}_{1,-\sqrt{3}}, {\hat{E}}_{-2,0}] = \sqrt{2} {\hat{E}}_{-1,-\sqrt{3}}. \label{eq:a-3}
\end{split}
\end{equation} 
In Eq.~(\ref{eq:a-2}), $c_{-1,\sqrt{3}} = c_{2,0} = 0$, since ${\hat{E}}_+$ and ${\hat{E}}_-$ are the raising and lowering operators and ${\hat{J}}_z$ and $\hat{Y}$ are 
linearly independent of ${\hat{E}}_{\bm{\alpha}}$'s. 
Therefore,  
\begin{align}
	&{\hat{E}}_+ = {\hat{E}}_{1,\sqrt{3}} + c_{1,-\sqrt{3}} {\hat{E}}_{1,-\sqrt{3}} + c_{-2,0} {\hat{E}}_{-2,0}, \label{eq:a-4} \\
	&[ {\hat{E}}_+, {\hat{E}}_- ] = ( 1 - 2 c_{-2,0}^2 + c_{1,-\sqrt{3}}^2 ) {\hat{J}}_z + \sqrt{3} ( 1 + c_{1,-\sqrt{3}}^2 ) \hat{Y}, \label{eq:a-5} 
\end{align}
\begin{equation}
\begin{split}
	&[ ( 1 - 2 c_{-2,0}^2 + c_{1,-\sqrt{3}}^2 ) {\hat{J}}_z + \sqrt{3} ( 1 + c_{1,-\sqrt{3}}^2 ) \hat{Y}, {\hat{E}}_+ ] \\
	= &2 \Bigl [ ( 2 - c_{-2,0}^2 + 2 c_{1,-\sqrt{3}}^2 ) {\hat{E}}_{1,\sqrt{3}} - c_{1,-\sqrt{3}} ( 1 + c_{-2,0}^2 + c_{1,-\sqrt{3}}^2 ) {\hat{E}}_{1,-\sqrt{3}} \\ 
	& - c_{-2,0} ( 1 - 2 c_{-2,0}^2 + c_{1,-\sqrt{3}}^2 ) {\hat{E}}_{-2,0} \Bigr ]. \label{eq:a-6} 
\end{split}
\end{equation} 
Here, the right-hand side of Eq.~(\ref{eq:a-6}) should be proportional to ${\hat{E}}_+$, which is satisfied if and only if 
$(c_{1,-\sqrt{3}}, c_{-2,0}) = (\pm 1, 0)$ or $(0,0)$, where the former case corresponds to type 1 and the latter one to type 2. 
Hence, we can conclude that there are only two classes of su(2) subalgebras with different magnitudes of the structure constants. 


\begin{thebibliography}{99} 
%Squeezing 
\bibitem{KU} M. Kitagawa and M. Ueda, Phys. Rev. A{\bf 47}, 5138 (1993) %interferometry
\bibitem{Wine}  D. J. Wineland, J. J. Bollinger, W. M. Itano, and F. L. Moore, Phys. Rev. A{\bf 46} R6797 (1992); D. J. Wineland, J. J. Bollinger, and W. M. Itano, Phys. Rev. A{\bf 50}, 67 (1994) %interferometry 
% general measurement 
%Wine
\bibitem{Maccone} V. Giovannetti, S. Lloyd, L. Maccone, Science {\bf 306}, 1330 (2004) %measurement 
\bibitem{DalRev} B. J. Dalton and S. Ghanbari, J. Mod. Opt. {\bf 59}, 287 (2011) %josephson (2-sites, 2-component) 
\bibitem{Nori} J. Ma, X Wanga, C. P. Suna, F. Nori, Phys. Rep. {\bf 509}, 89 (2011) %Review 
\bibitem{Klempt} B. L\"cke,  M. Scherer,  J. Kruse, L. Pezz\'e, F. Deuretzbacher, P. Hyllus, O. Topic, J. Peise, W. Ertmer, J. Arlt, L. Santos, A. Smerzi, C. Klempt, Science {\bf 334}, 773 (2011) %two-mode squeezing 
\bibitem{GrossRev} C. Gross, J. Phys. B: At. Mol. Opt. Phys. {\bf 45}, 103001 (2012) %2-component BEC 
%magnetometer 
\bibitem{Mabuchi1} J. M. Geremia, J. K. Stockton, A. C. Doherty, and H. Mabuchi, Phys. Rev. Lett. {\bf 91}, 250801 (2003); J. M. Geremia, J. K. Stockton, and H. Mabuchi, Science 
{\bf 304}, 270 (2004) 
\bibitem{Mabuchi2} J. M. Geremia, J. K. Stockton, and H. Mabuchi, Phys. Rev. Lett. {\bf 94}, 203002 (2005); \textit{ibid.}, Phys. Rev. A{\bf 73}, 042112 (2006) %generate squeezing 
\bibitem{Molmer1} V. Petersen, L. B. Madsen, and K. M{\o}lmer, Phys. Rev. A {\bf 71}, 012312 (2005) %magnetometer
\bibitem{Stamper-Kurn1} M. Vengalattore, J. M. Higbie, S. R. Leslie, J. Guzman, L. E. Sadler, and D. M. Stamper-Kurn, Phys. Rev. Lett. {\bf 98}, 200801 (2007) 
\bibitem{Stamper-Kurn2} J. D. Sau, S. R. Leslie, M. L. Cohen, and D. M. Stamper-Kurn, New. J. Phys. {\bf 12}, 085011 (2010) %nematic squeezing 
\bibitem{Thompson} Z. Chen, J. G. Bohnet, S. R. Sankar, J. Dai, and J. K. Thompson, Phys. Rev. Lett. {\bf 106}, 133601 (2011) %cavity QED 
%optical lattice clock 
%Wine
\bibitem{Polzik1} D. Oblak, P. G. Petrov, C. L. G. Alzar, W. Tittel, A. K. Vershovski, J. K. Mikkelsen, J. L. S{\o}rensen, and E. S. Polzik, 
Phys. Rev. A {\bf 71}, 043807 (2005) 
\bibitem{Oates} A. D. Ludlow, T. Zelevinsky, G. K. Campbell, S. Blatt, M. M. Boyd, M. H. G. de Miranda, M. J. Martin, J. W. Thomsen, S. M. Foreman, Jun Ye, T. M. Fortier, J. E. Stalnaker, S. A. Diddams, Y. Le Coq, Z. W. Barber, N. Poli, N. D. Lemke, K. M. Beck, C. W. Oates, Science {\bf 319}, 1805 (2008) 
\bibitem{Ye} M. J. Martin, M. Bishof, M. D. Swallows, X. Zhang, C. Benko, J. von-Stecher, A. V. Gorshkov, A. M. Rey, and Jun Ye, arXiv:1212.6291 (2012) 
%realization of squeezed spin state 
\bibitem{Polzik3} J. Hald, J. L. S{\o}rensen, C. Schori, and E. S. Polzik, Phys. Rev. Lett. {\bf 83}, 1319 (1999) 
\bibitem{Bigelow} A. Kuxmich, L. Mandel, J. Janis, Y. E. Young, R. Ejnisman, and N. P. Bigelow, Phys. Rev. A{\bf 60}, 2346 (1999); A. Kuzmich, L. Mandel, and N. P. Bigelow, Phys. Rev. Lett. {\bf 85}, 1594 (2000) %QND 
\bibitem{Kasevich} C. Orzel, A. K. Tuchman, M. L. Fenselau, M. Yasuda, M. A. Kasevich, Science {\bf 291}, 2386 (2001) 
\bibitem{Polzik4} B. Julsgaard, A. Kozhekin, and E. S. Polzik, Nature, {\bf 413}, 400 (2001) 
\bibitem{Oberthaler1} J. Est\`eve, C. Gross, A. Weller, S. Giovanazzi, and M. K. Oberthaler, Nature {\bf 455}, 1216 (2008) %two sites 
\bibitem{Polzik5} J. Appel, P. J. Windpassinger, D. Oblak, U. B. Hoff, N. Kj{\ae}rgaard, and E. S. Polzik, Proc. Nat. Acad. Sci. U.S.A. {\bf 106}, 10960 (2009) 
\bibitem{Vuletic} I. D. Leroux, M. H. Schleier-Smith, and V. Vuleti\'c, Phys. Rev. Lett. {\bf 104}, 073602 (2010); Monika H. Schleier-Smith, Ian D. Leroux, and Vladan Vuleti\'c, Phys. Rev. A{\bf 81}, R021804 (2010). %atom-field, one-axis 
\bibitem{Oberthaler2} C. Gross, T. Zibold, E. Nicklas, J. Est\`eve, and M. K. Oberthaler, Nature {\bf 464}, 1165 (2010) % two levels, below SQL 
\bibitem{Treutlein} M. F. Riedel, P. B\"ohi, Y. Li, T. W. H\"ansch, A. Sinatra, and P. Treutlein, Nature {\bf 464}, 1170 (2010) % two levels, below SQL 
%Thompson
\bibitem{Chapman} C. D. Hamley, C. S. Gerving, T. M. Hoang, E. M. Bookjans and M. S. Chapman, Nature Phys. {\bf 8}, 305 (2012) 
\bibitem{Takahashi2} R. Inoue, S. Tanaka, R. Namiki, T. Sagawa, and Y. Takahashi, arXiv:1301.1016 (2013) 
%coherent spin state 
\bibitem{Radcliffe} J. M. Radcliffe, J. Phys. A: Gen. Phys. {\bf 4}, 313 (1971). 
\bibitem{Arecchi} F. T. Arecchi, E. Courtens, R. Gilmore, and H. Thomas, Phys. Rev. A {\bf 6}, 2211 (1972). 
%two components BEC  
%DalRev 2 sites, 2 levels 
%GrossRev 2 sites 
\bibitem{Molmer2} A. S{\o}rensen and K. M{\o}lmer, Phys. Rev. Lett. {\bf 83}, 2274 (1999) %two sites
\bibitem{Zoller1} A. S{\o}rensen, L. -M. Duan, J. I. Cirac, and P. Zoller, Nature {\bf 409}, 63 (2001) %two levels, one-axis, atom-atom 
%generating SSS
%imprint 
%nonlinear interaction 
%DalRev (2011) atom-atom 
%GrossRev (2012) atom-atom 
%Mabuchi atom-field 
%Tompson (2011) atom-field 
%Zoller atom-atom 
%Oberthaler1
%Obsethaler2
%Treutlein 
\bibitem{Collet} M. J. Steel and M. J. Collet, Phys. Rev. A {\bf 57}, 2920 (1998) %atom atom josephson 
\bibitem{Lukin} A. Andr\'e, L.-M. Duan, and M. D. Lukin, Phys. Rev. Lett. {\bf 88}, 243602 (2002) 
\bibitem{Takahashi1} M. Takeuchi, S. Ichihara, T. Takano, M. Kumakura, T. Yabuzaki, and Y. Takahashi, Phys. Rev. Lett. {\bf 94}, 023003 (2005) 
\bibitem{LiYou1} Y. C. Liu, Z. F. Xu, G. R. Jin, and L. You, Phys. Rev. Lett. {\bf 107}, 013601 (2011) 
%nematic tensor 
%
%spin-nematic squeezing 
\bibitem{Polzik2} A. Kuzmich, Klaus M{\o}lmer, and E. S. Polzik, Phys. Rev. Lett. {\bf 79}, 4782 (1997); J. L. S{\o}rensen, J. Hald, and E. S. Polzik, Phys. Rev. Lett. {\bf 80}, 3487 (1998) 
%Stamper-Kurn2 
%two-mode squeezing 
\bibitem{Meystre} H. Pu and P. Meystre, Phys. Rev. Lett. {\bf 85}, 3987 (2000) 
\bibitem{Zoller2} L.-M. Duan, A. S{\o}rensen, J. I. Cirac, and P. Zoller, Phys. Rev. Lett. {\bf 85}, 3991 (2000) 
\bibitem{LiYou2} \"O. E. M\"ustecapl{\i}o\u{g}lu, M. Zhang, and L. You, Phys. Rev. A{\bf 66}, 033611 (2002) 
\end{thebibliography}

\end{document}